\documentclass[aps,twocolumn,showpacs,prl,amsmath,amssymb,floatfix,superscriptaddress]{revtex4-2}
\usepackage{color}
\usepackage[usenames,dvipsnames]{xcolor}
\usepackage[colorlinks=true,citecolor=Cerulean,linkcolor=RubineRed,urlcolor=Cerulean]{hyperref}
\usepackage[normalem]{ulem}
\usepackage{soul,xcolor}
\usepackage{dcolumn}
\usepackage{bm}
\usepackage{amsmath}
\usepackage{amssymb}
\usepackage{amsthm}
\usepackage{mathtools}
\usepackage{ulem}
\usepackage{graphicx}
\usepackage{epsfig}
\usepackage{epstopdf}
\usepackage{fontenc}
\usepackage{tipa}
\usepackage{url}
\usepackage{booktabs}
\usepackage{multirow}
\usepackage{float}
\usepackage{subcaption}
\usepackage{ragged2e}
\begin{document}

\title{Thermodynamic Geometric Constraint on the Spectrum of Markov Rate Matrices}
\author{Guo-Hua Xu}
\affiliation{Universal Biology Institute, The University of Tokyo, 7-3-1 Hongo, Bunkyo-ku, Tokyo 113-0033, Japan}
\email{guohua.xu@ubi.s.u-tokyo.ac.jp}

\author{Artemy Kolchinsky}
\affiliation{ICREA-Complex Systems Lab, Universitat Pompeu Fabra, 08003 Barcelona, Spain}

\author{Jean-Charles Delvenne}
\affiliation{ICTEAM Institute, UCLouvain, Louvain-la-Neuve, Belgium}
\affiliation{Department of Physics, Kyoto University, Japan}

\author{Sosuke Ito}
\affiliation{Universal Biology Institute, The University of Tokyo, 7-3-1 Hongo, Bunkyo-ku, Tokyo 113-0033, Japan}
\affiliation{Department of Physics, The University of Tokyo, 7-3-1 Hongo, Bunkyo-ku, Tokyo 113-0033, Japan}

\begin{abstract}
    
The spectrum of Markov generators encodes physical information beyond simple decay and oscillation, which reflects irreversibility and governs the structure of correlation functions. 
In this work, we prove an ellipse theorem that provides a universal thermodynamic geometric constraint on the spectrum of Markov rate matrices. 
The theorem states that all eigenvalues lie within a specific ellipse in the complex plane. 
In particular, the imaginary parts of the spectrum, which indicate oscillatory modes, are bounded by the maximum thermodynamic force associated with individual transitions.
This spectral bound further constrains the initial short-time behavior of correlation functions between two arbitrary observables.
Finally, we compare our result with a previously proposed conjecture, which remains an open problem and warrants further investigation.

\end{abstract}

\maketitle

The spectral properties of linear operators are fundamental to the analysis of physical systems. 
In open classical and quantum systems, the eigenvalues of the dynamical generator directly govern relaxation and oscillations \cite{van1992stochastic,breuer2002theory,rivas2012open}: The real parts of the eigenvalues, being negative, govern the exponential decay of the eigenmodes, whereas the imaginary parts encode the frequencies of the oscillatory modes. 

A particularly important class of generators is the rate matrix of a continuous-time Markov jump process. In graph theory, such matrices appear as the Laplacian matrices of weighted graphs \cite{Merris_Laplacian_1994,chung1997spectral,Schnakenberg_Network_1976}. 
Rate matrices play a crucial role in the stochastic thermodynamics of discrete-state systems \cite{Seifert_2012,Review_modernthermodynamics,Barato_originTUR_2015,Gingrich_proveTUR_2016,Dechant_corrTUR_2021,Vo_KTUR_2022,Shiraishi_SpeedLimit_2018,Lee_SpeedLimit_2022,Ito_infogeo_2018,dechant2022minimum, yoshimura2023housekeeping, Tan_infogeo_2023,Bagrets_counting_2003,Sinitsyn_2007,Ren_counting_2010,Owen_NoneqResponse_2020,Esposito_NoneqResponse_2024,Esposito_NoneqResponse_2025,kwon2024fluctuation,Harunari_MutualLinearity_2024,cengio2025mutual,Khodabandehlou_2025,bao2024nonequilibrium,bauer2025stroboscopicmeasurementsmarkovnetworks} and find broad applications in complex networks \cite{Boccaletti_complex_2006,Masuda_random_2017}, chemical reaction networks\cite{Gaspard_corr_2002,Gaspard_Trace_2002}, biology\cite{Munsky_gene_2009,barkai1997robustness,Tu_ultrasensitivity_2008}, and economics\cite{Jarrow_ReviewFinancial_1997,rolski2009stochastic}. 

The spectrum of Markov generators encodes rich physical information that goes beyond simple decay rates and oscillatory behavior \cite{Minganti_SpectralGKLS_2018,Gaveau_metastab_1998,Mori_SymmetrizedGap_2023,Sawada_Topology_2024,Kawabata_LSM_2024}. It captures the degree of irreversibility in the system \cite{Uhl_conjecture_2019,Barato_cohcyc_2017,Clara_bioosc_2020,Oberreiter_cohcyc_2022,Remlein_cohcyc_2022,santolin_cohcyc_2025,Kolchinsky_SpectralBound_2024} and governs the structure of correlation functions \cite{Ohga_corsscorr_2023,Shiraishi_boundfluc_2023,Bakewell-Smith_iTUR_2023,Tuan_iTUR_2025, aguilera2025inferring}.
As an extension of the conjecture proposed only for the second eigenvalue \cite{Barato_cohcyc_2017}, recent work by Ohga {\it et al.}~\cite{Ohga_corsscorr_2023} shows that the spectrum exactly lies within a sector in the complex plane, with the opening angle determined by the maximal thermodynamic force of a cycle. This spectral constraint also follows from a thermodynamic bound on cross-correlation functions. A key insight is that oscillatory modes and asymmetric correlations appear only in nonequilibrium systems.

While the sector bound captures certain aspects of irreversibility, a deeper geometric understanding of the spectrum remains elusive. This naturally raises two fundamental questions:
(i) Can we identify a universal constraint on the geometry of the spectrum that arises from irreversibility?
(ii) How is this spectral geometry quantitatively connected to the (initial) time derivative of correlation functions?

\begin{figure}
\centering
    \includegraphics[width=1 \columnwidth]{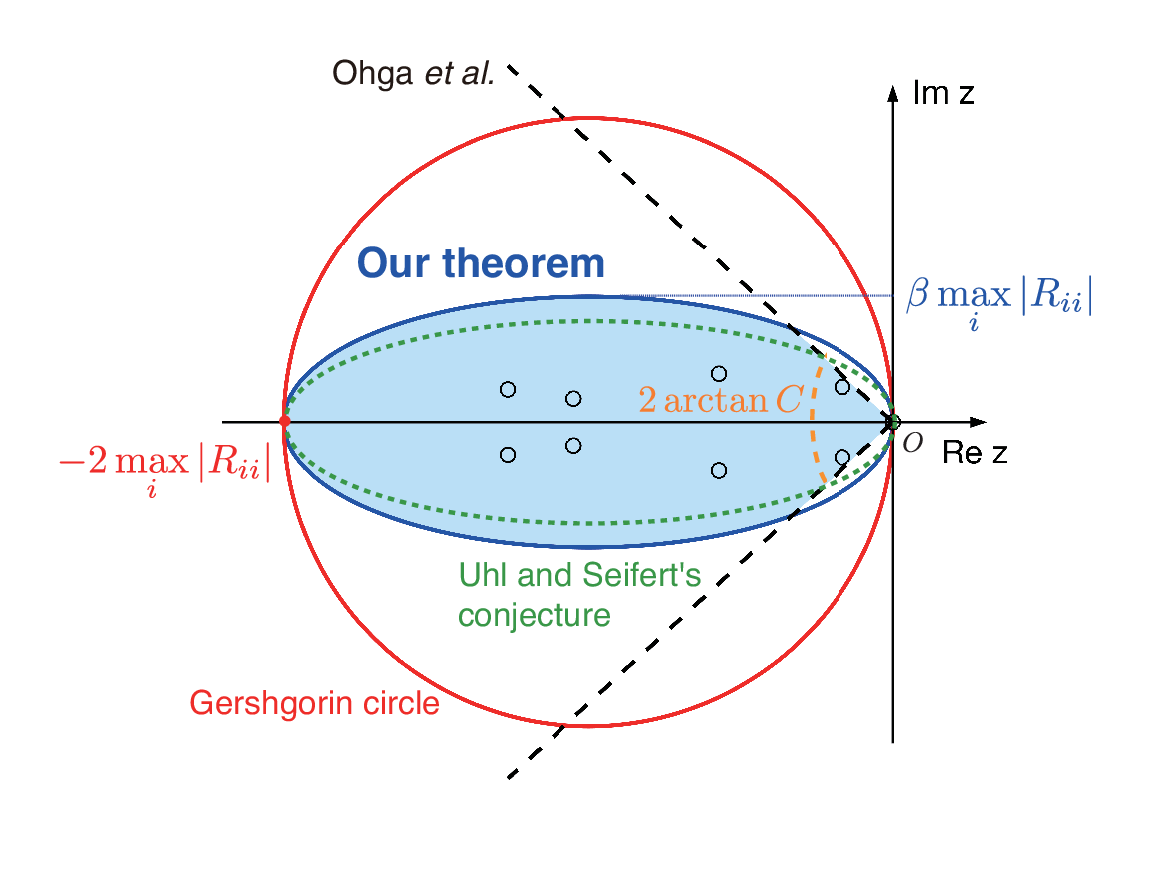}
    \caption{\RaggedRight Bounds on the spectrum of Markov rate matrices. The blue line shows the elliptical bound derived in this work, and the open circles denote the eigenvalues.
    Our theorem states that the spectrum is contained within this ellipse. The major axis lies along the real axis and spans the interval from $-2 \max_i |R_{ii}|$ to $0$, where $|R_{ii}|$ is the escape rate. The minor axis has length $2 \beta \max_i |R_{ii}|$ with $\beta= \tanh \left(\max_e F_e /2\right)$, where $F_e$ denotes the thermodynamic force on edge $e$.
    The black dashed lines correspond to the sectorial bound reported by Ohga {\it et al.} in Ref.~\cite{Ohga_corsscorr_2023}. 
    The opening angle of the sector is bounded above by $2 \arctan{C}$, where $C = \max_c \{ \tanh[\mathcal{F}_c/(2 n_c)]/ \tan(\pi/n_c) \} \leq \max_c \mathcal{F}_c/(2 \pi)$ with $\mathcal{F}_c$ and $n_c$ denoting the thermodynamic force and the length of cycle $c$, respectively. Therefore, the spectrum is restricted to the blue shaded region. 
    The red circle represents the Gershgorin circle bound, and the green dashed ellipse represents the bound numerically conjectured by Uhl and Seifert~\cite{Uhl_conjecture_2019}. Here, the vertical extents of the Gershgorin circle and the ellipse conjectured by Uhl and Seifert are $2 \max_i |R_{ii}|$ and $2 \beta' \max_i |R_{ii}|$, respectively, where $\beta' = \tanh \left( \max_c \{ \mathcal{F}_c /(2n_c) \}\right)$.
    }
    \label{fig:ellipse}
\end{figure}

In this Letter, we establish a unified framework addressing the above questions by revealing the intrinsic connection between the spectrum and correlation functions through the mathematical concept of the numerical range. Inspired by Uhl and Seifert’s conjecture \cite{Uhl_conjecture_2019}, we derive an ellipse theorem that provides a universal thermodynamic bound on the spectrum of Markov rate matrices. Our theorem states that the spectrum lies within a specific ellipse in the complex plane, as illustrated in Fig.~\ref{fig:ellipse}. The vertical semi-axis of the ellipse depends on the degree of irreversibility, which vanishes for the detailed balance condition and reaches the circular limit in the case of infinite irreversibility. 

Our result reveals a fundamental thermodynamic constraint on the spectrum of Markov generators that goes beyond the Gershgorin circle theorem \cite{gershgorin1931uber,weisstein2003gershgorin}. 
The Gershgorin circle theorem is a well-known result in matrix analysis that bounds the location of eigenvalues within certain disks determined purely by the matrix entries, independent of any physical considerations. 
Unlike this purely algebraic and general bound, our ellipse incorporates thermodynamic considerations and directly reflects the irreversibility inherent to nonequilibrium dynamics. This connection indicates that the mathematical structure of the spectrum is not merely a consequence of abstract algebraic properties but is fundamentally shaped and constrained by physical laws.

\textit{Numerical range and correlation functions.---}
Mathematically, the rate matrix (or generator) $R \in \mathbb{R}^{N \times N}$ of a continuous-time Markov process on a finite state space of size $N$ satisfies $R_{ij} \geq 0$ for $i \neq j$ and $\sum_i R_{ij} = 0$ for all $j$. 
We consider a rate matrix $R$ with a strictly positive steady state $\mathbf{\pi}$, satisfying $R \mathbf{\pi} = 0$
\footnote{
    Our result still holds when $R$ is a reducible matrix where the steady-state distribution is not unique, as long as one can choose a steady-state distribution with strictly positive entries. 
}.

We introduce the matrix \cite{Mori_SymmetrizedGap_2023,Kolchinsky_SpectralBound_2024}
\begin{equation}
\tilde{R} \coloneqq \Pi^{-1/2} R \,\Pi^{1/2},
\end{equation}
where $\Pi = {\rm diag} (\pi_1, \pi_2, \dots, \pi_N)$.
The numerical range $W(\tilde{R})$ of the matrix $\tilde{R}$ is defined as the set of complex numbers \cite{axler1997numerical},
\begin{equation}
W(\tilde{R}) = \left\{ \frac{ \langle \mathbf {x}|\tilde{R}|\mathbf {x} \rangle}{\langle \mathbf {x} | \mathbf {x} \rangle} \Bigg| ~\mathbf {x} \in \mathbb{C}^n ,\mathbf {x} \neq 0 \right\}.
\end{equation}
By introducing $\mathbf {x}=\Pi^{1/2}(\mathbf {a}+i\mathbf {b})$, where $\mathbf {a}$ and $\mathbf {b}$ are two arbitrary state observables, we show that
$W(\tilde{R})$ is given by the correlation functions:
\begin{align}
\frac{ \langle \mathbf {x}|\tilde{R}|\mathbf {x} \rangle}{\langle \mathbf {x} | \mathbf {x} \rangle} 
&= \frac{\mathbf {a}^{\top} R\Pi\, \mathbf {a} + \mathbf {b}^{\top} R\Pi\, \mathbf {b} + i (\mathbf {a}^{\top} R\Pi\, \mathbf {b} - \mathbf {b}^{\top} R\Pi\, \mathbf {a})}{\mathbf {a}^{\top}\Pi\, \mathbf {a} + \mathbf {b}^{\top}\Pi\, \mathbf {b}} \notag \\
&= \frac{\dot{C}_{aa}(0)+\dot{C}_{bb}(0) + i (\dot{C}_{ab}(0)-\dot{C}_{ba}(0))}{C_{aa}(0) + C_{bb}(0)} , \label{NR_corr}
\end{align}
where $C_{ab}(t) \coloneqq \mathbf {a}^{\top}e^{Rt} \Pi \mathbf {b}$, and $\dot{C}_{ab}(t)$ denotes the time derivative. 
Interestingly, $W(\tilde{R})$ encodes the correlation structure: its imaginary part captures the asymmetry in cross-correlations, while its real part reflects the decay of auto-correlations.
Thus, our results provide a precise constraint on the short-time behavior of $C_{ab}$. For finite times $t > 0$, bounds on the decay of autocorrelation functions can be obtained via a numerical range interpretation. We return to this point at the end.

Note that the eigenvalues of $R$ (or $\tilde{R}$) can be obtained by considering the specific observables $\mathbf {a}$ and $\mathbf {b}$ as discussed in Ref. \cite{Ohga_corsscorr_2023} (see the {\it End Matter}). 
These observables correspond to special cases where the vector $\mathbf {x}$ becomes an eigenvector of $\tilde{R}$.
By definition, the numerical range satisfies $\sigma(\tilde{R}) \subset W(\tilde{R})$, where $\sigma(\tilde{R})$ is the spectrum of the matrix $\tilde{R}$ defined as
\begin{equation} \notag
\sigma(\tilde{R}) = \left\{ \frac{ \langle \tilde{ \mathbf{u} }|\tilde{R}|\tilde{ \mathbf{u} } \rangle}{\langle \tilde{ \mathbf{u} } | \tilde{ \mathbf{u} } \rangle} \Bigg| \, \exists \lambda \in \mathbb{C}, \tilde{R} \tilde{ \mathbf{u} } = \lambda \tilde{ \mathbf{u} }, \tilde{ \mathbf{u} } \in \mathbb{C}^n, \tilde{ \mathbf{u} } \neq 0 \right\}.
\end{equation}
Since $\tilde{R}$ is defined as a similarity transformation of $R$, they have the same spectra $\sigma(R)=\sigma(\tilde{R})$, and therefore $\sigma(R) \subset W(\tilde{R})$. 
Thus, the numerical range $W(\tilde{R})$ bridges the spectral properties of the rate matrix $R$ with the possible values of the correlation functions.


\textit{Ellipse theorem.---}
The numerical range $W(\tilde{R})$ is contained within an ellipse $EL$ in the complex plane, i.e., $ W(\tilde{R}) \subset EL$, where
\begin{equation}
EL \coloneqq \left\{z\in \mathbb{C} \Bigg| \left( \frac{\mathrm{Re}\,z + \alpha}{\alpha} \right)^2 + 
\left( \frac{\mathrm{Im}\,z}{\alpha \beta} \right)^2
\leq 1 \right\}. \notag
\end{equation}
Here,
$
\alpha = \max_i |R_{ii}|
$
denotes the largest escape rate from a state, 
$
\beta = \tanh \left(\max_e F_e /2\right) < 1
$
quantifies the degree of irreversibility,
and $F_e$ is the thermodynamic force associated with the transition (edge) $e=j \to i$:
\begin{equation}
    F_e = \ln \frac{R_{ij} \pi_j}{R_{ji} \pi_i} .
\end{equation}
The maximum in the definition of $\beta$ is taken over all edges with $R_{ij} > 0$ and $R_{ji} > 0$; 
if there exists any edge with a finite rate in one direction and zero in the other, we set $\beta = 1$.

A direct consequence of the ellipse theorem is that the spectrum of $R$ is constrained within the ellipse, $\sigma(R) \subset EL$, because $\sigma(R) = \sigma(\tilde{R}) \subset W(\tilde{R})$.
Physically, the ellipse theorem implies that the emergence of complex eigenvalues, which indicate oscillations, and asymmetric correlations
is fundamentally limited by the degree of irreversibility. 

Consider a complex eigenvalue $\lambda_n = {\rm Re} \lambda_n + i\, {\rm Im} \lambda_n$. The corresponding eigenmode $\mathbf{u}_n$, satisfying $R \mathbf{u}_n = \lambda_n \mathbf{u}_n$, contributes to the time evolution of the probability distribution. Suppose that the initial state is given by $\mathbf{p} (0)= \sum_n k_n(0) \, \mathbf{u}_n$ where $k_n(0)$ are determined by the initial condition. Since the dynamics of the continuous-time Markov processes follows $\mathrm{d} \mathbf{p}/\mathrm{d}t =R \mathbf{p}$, the time evolution of the probability is given by $\mathbf{p}(t) = \sum_n k_n(t) \mathbf{u}_n$ with $k_n (t) =\exp [{ ( {\rm Re} \lambda_n) t} + {i( {\rm Im} \lambda_n)t }]\, k_n (0)$
\footnote{
For diagonalizable $R$, we have left eigenvectors $\mathbf{v}_n$ satisfying $ \mathbf{v}_n^{\dagger} R= \mathbf{v}_n^{\dagger} \lambda_n$, and the coefficients for the initial condition are given by $k_n(0) = \mathbf{v}_n^{\dagger} \, \mathbf{p}(0)$. However, if $R$ is not diagonalizable, the initial state may involve generalized eigenvectors (root vectors), which lead to additional dynamics of the form $t^k e^{\lambda t}$. In this case, we restrict our analysis to eigenmodes and neglect such contributions arising from nontrivial Jordan blocks.
}. 
Here, the imaginary part ${\rm Im} \lambda_n$ determines the frequency of the oscillation, $|{\rm Im} \lambda_n| = \omega_n$.
Consequently, the frequency of any oscillation mode in a Markov jump process is bounded by the maximum thermodynamic force in the system:
\begin{equation} 
\max_n \omega_n \leq \alpha \tanh \left(\max_e F_e /2\right) , \label{frequency}
\end{equation}
which is indicated by the dashed blue horizontal line in Fig.~\ref{fig:ellipse}.  
This inequality provides a thermodynamic bound on the generation of nontrivial temporal structures in stochastic dynamics. Remarkably, it implies that not only the slowest-decaying modes, but also the fast-decaying ones cannot exhibit strong oscillations unless $\beta$ is large. This is somewhat surprising, as thermodynamic constraints are typically associated with slowly-decaying oscillatory modes, rather than those that decay rapidly \cite{Barato_cohcyc_2017,Clara_bioosc_2020,Oberreiter_cohcyc_2022,Remlein_cohcyc_2022,santolin_cohcyc_2025,Ohga_corsscorr_2023}.
Importantly, this result provides a fundamental constraint on biochemical clocks \cite{Barato_cohcyc_2017,Clara_bioosc_2020,Oberreiter_cohcyc_2022,cao2015bio_osc,zheng2024topological}, such as circadian rhythms \cite{Nakajima_KaiABC_2005,partch2014molecular,takahashi2017transcriptional}, where nonequilibrium oscillations are essential for timekeeping in living systems. 

\textit{Sketch of the proof.---}
We provide two different derivations, a mathematically oriented one is presented in Supplementary Material \cite{supp}, while the main text and the \textit{End Matter} focus on a physically motivated proof, with the full details given in the \textit{End Matter}.
To prove the ellipse theorem, we decompose the matrix $\tilde{R}$ into its symmetric and antisymmetric parts:
$\mathrm{Sym}(\tilde{R}) \coloneqq (\tilde{R} + \tilde{R}^\top)/2$ and
$\mathrm{Skew}(\tilde{R}) \coloneqq (\tilde{R} - \tilde{R}^\top)/2$,
which correspond to the real and imaginary parts of the numerical range $W(\tilde{R})$, respectively.
These components relate to physical quantities, where the antisymmetric part provides the current matrix $J=2\,\Pi^{\frac{1}{2}}\,\mathrm{Skew}(\tilde{R})\,\Pi^{\frac{1}{2}}$ and the symmetric part provides the traffic matrix $A=2 \, \Pi^{\frac{1}{2}} \, \mathrm{Sym}(\tilde{R}) \, \Pi^{\frac{1}{2}}$.  
From Eq.~\eqref{NR_corr}, the imaginary part $y := {\rm Im}z$ of $z \in W(\tilde{R})$ can be written as $\sum_{ij} J_{ij} (a_i b_j - a_j b_i)$ over a normalizing factor. Similarly, the real part $x:= {\rm Re}z$ is given by $\sum_{ij} A_{ij} (a_i a_j + b_j b_i)$ over a normalizing factor. 
Using $\tanh (F_e/2)= {J_{ij}}/{A_{ij}}$ with $e = j \to i$, we scale $y$ and define $\tilde{y} \coloneqq \sum_{ij} A_{ij} |a_i b_j - a_j b_i|$, so that $\tilde{y} \geq |y|/\beta$ with $\beta \coloneqq \tanh \left(\max_e F_e /2\right) $. 
Then, we decompose $x$ as $x = x_{\rm D} + x_{\rm ND}$, where $x_{\rm D}$ collects the diagonal contributions of $A$ and $x_{\rm ND}$ the off-diagonal parts. By construction, $\tilde{y}$ and $x_{\rm ND}$ involve the same off-diagonal index pairs on which the summands are nonzero. As shown in the \textit{End Matter}, this yields $x_{\rm ND}^2 + \tilde{y}^2 \leq x_{\rm D}^2$, i.e., $(x-x_{\rm D})^2 + \tilde{y}^2 \leq x_{\rm D}^2$, which is a circular bound. Since $-\alpha< x_{\rm D} <0$ and $\tilde{y} \geq |y|/\beta$, we obtain the elliptic bound.


\begin{figure}
    \centering
    \includegraphics[width=1 \columnwidth]{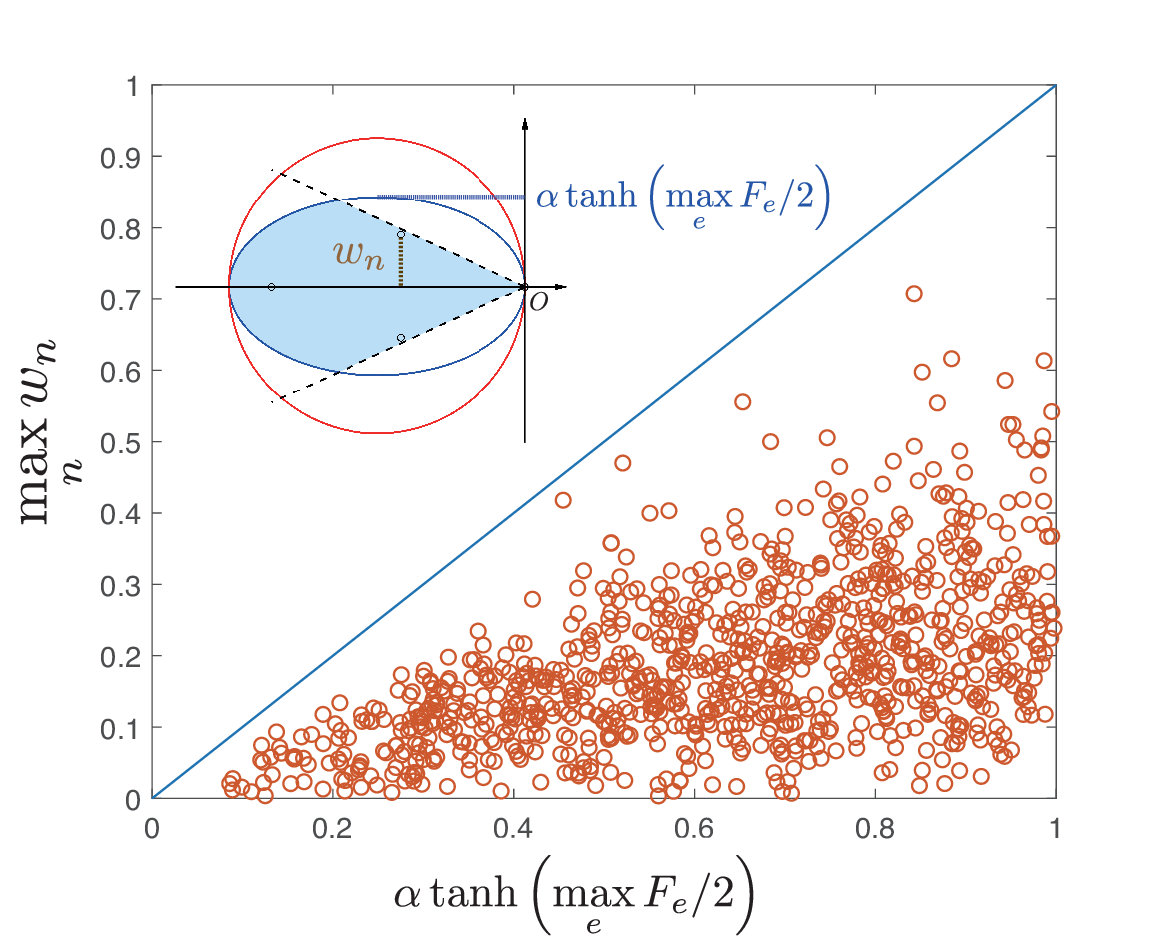}
    \caption{\RaggedRight Thermodynamic bound on maximal frequency for cycles with $N=4$ states. 
    The open circles represent the maximum imaginary part of the eigenvalues of rate matrices corresponding to random walks on a ring with random jump rates. The blue line represents the bound given by Eq.~\eqref{frequency}. }
    \label{fig:example}
\end{figure}

\textit{Markov jump process on a single cycle.---} 
As an example, we consider the random walk on a ring with $N$ sites. First, we discuss the case of the uniform cycle in which the forward jump rate $\omega_+$ and the backward jump rate $\omega_-$ are constant. In this case, the eigenvalues can be calculated analytically \cite{Uhl_conjecture_2019}, $\lambda_n = \alpha[-1+\cos(2\pi N/n) + i \tanh (F_e/2) \sin (2\pi N/n)]$, where $\alpha = \omega_+ + \omega_-$ and $F_e = \ln(\omega_+/\omega_-)$. The spectrum is located precisely on the boundary of the ellipse. 
This example confirms that the spectral bound given by the ellipse theorem is tight, as it is exactly saturated by the uniform cycle.

Then, we consider the random walk on a ring with different jump rates, a minimal model with applications in physics \cite{Sawada_Topology_2024,Seifert_2012}, and biology \cite{Owen_NoneqResponse_2020,Mehta_bio_2012,Bialek_bio_2005,li2003sensitivity}.
We simulate jump rates as independent random variables uniformly distributed in the interval $(0,1)$.
For each realization, we compute the maximum imaginary part of the eigenvalues of the rate matrix.
The thermodynamic bound on the maximal frequency, given by inequality~\eqref{frequency}, is numerically validated in Fig.~\ref{fig:example}. This confirms that the oscillatory modes in randomly disordered cycles remain constrained by the thermodynamic force, as predicted by our theory.
Although we have demonstrated the result for a single-cycle example, our theoretical framework holds for Markov processes with arbitrary network topologies.

\begin{figure}
    \centering
    \includegraphics[width=1 \columnwidth]{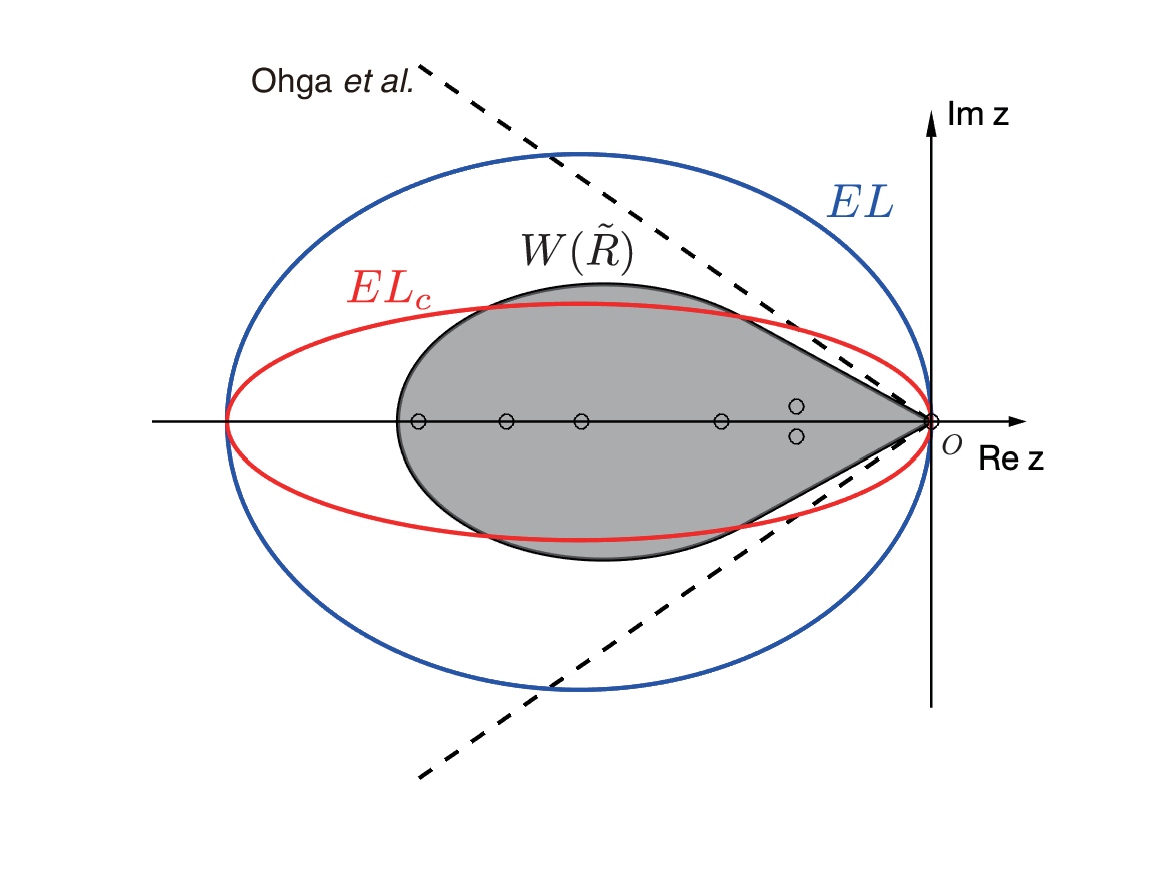}
    \caption{\RaggedRight
Comparison between the conjectured ellipse  $EL_c$  (red line) and the derived bound  $EL $ (blue line).
This example illustrates that the conjectured ellipse $EL_c$ contains all eigenvalues of $R$ (hollow circles), but does not enclose the numerical range  $W(\tilde{R})$  (black shaded region). The bound by Ohga {\it et al.} also holds for the numerical range $W(\tilde{R})$.
The numerical range is calculated using the algorithm in Ref. \cite{Johnson_calculation_1978}.
This example is obtained from a random walk on a ring (as in Fig.~2) with i.i.d. jump rates sampled uniformly from the interval $(0,1)$. We sample such random walks for increasing ring size $N$. For $N \geq 6$, cases where $EL_c$ does not enclose $W(\tilde{R})$ are observed, and the frequency of such cases increases with $N$. This figure shows a representative instance with $N = 7$, found within 1000 random samples.
}
    \label{fig:conjecture}
\end{figure}

\textit{Comparison with the conjectured bound.}---
The ellipse bound derived in this work differs from the previous conjecture \cite{Uhl_conjecture_2019} in the value of the minor axis. 
The conjecture assumes the existence of an ellipse $EL_c$ such that $\sigma(R) \subset EL_c$. 
The conjectured ellipse $EL_c$ shares the same major axis as our bound $EL$, but its semi-minor axis is given by $\alpha \tanh \left( \max_c \{ \mathcal{F}_c/(2n_c)\} \right) =: \alpha \beta'$, where $\mathcal{F}_c$ is the thermodynamic force along a cycle $c$, and $n_c$ is the number of states in that cycle. This expression uses the cycle-averaged force and then maximizes over all cycles. We observe that $\beta' \leq \beta$ due to the triangle inequality, i.e., $|\mathcal{F}_c|=|\sum_{e\in c} F_e| \leq \sum_{e\in c}|F_e|$, and the inequality between the maximum value and the arithmetic mean, i.e., $\max_{e \in c}|F_e| \geq \sum_{e\in c} {|F_e|/n_c} (\geq|\mathcal{F}_c|/n_c)$. Since the maximum in the definition of $\beta$ and $\beta'$ already implies positive $\mathcal{F}_c$ and $F_e$, it follows that $\max_{e}F_e \geq \max_{c} \mathcal{F}_c/n_c$.
Therefore, we can verify that $EL_c \subseteq EL$, meaning that the conjectured ellipse is strictly tighter.

In contrast, our ellipse theorem rigorously establishes that $\sigma(R) \subset W(\tilde{R}) \subset EL$, with the minor axis determined by the largest edge-wise thermodynamic force $F_e$. While the theorem $\sigma(R) \subset EL$ is a looser bound on the spectrum $\sigma(R)$ compared to the conjecture, $\sigma(R) \subset EL_c \subseteq EL$, our ellipse theorem may give a tight bound on the numerical range $W(\tilde{R})$. Indeed, our numerical simulations reveal that the bound defined by the conjectured ellipse $EL_c$ does not hold for the numerical range, $W(\tilde{R}) \not\subset EL_c$, while it still correctly bounds the eigenvalues, $\sigma(R) \subset EL_c$. 
An example demonstrating this discrepancy is shown in Fig.~\ref{fig:conjecture}. This indicates that the conjecture cannot be derived from the numerical range. Interestingly, the sectorial bound derived by Ohga {\it et al.}~\cite{Ohga_corsscorr_2023} based on cycle forces remains valid for the numerical range $W(\tilde{R})$, whereas the conjectured ellipse bound based on cycle forces is not. 
The validity of the conjectured bound remains an open question, warranting further theoretical and numerical investigation.

\textit{Conclusion and discussion.}---
In summary, we have established a thermodynamic geometric bound on the spectrum of Markov rate matrices by introducing the numerical range which bridges the spectrum and the correlation functions. The derived ellipse theorem reveals that all eigenvalues lie within a bounded elliptical region in the complex plane, with the geometry determined by the maximal escape rate and the maximal thermodynamic force. 
Our findings provide new insights into how irreversibility shapes the spectral structure and dynamical behavior in stochastic systems.

We also compared our result with a previously conjectured spectral bound \cite{Uhl_conjecture_2019}, which is tighter but remains an open problem. Further investigations based on the properties of the eigenvectors may help refine our results and eventually lead to a proof of the conjecture. Additionally, the numerical range may offer valuable insights into spectral properties in quantum Markov systems. However, the structure of quantum master equations is significantly more complex, requiring further study.

We briefly discuss how the spectral gap is introduced in our analysis and how it constrains autocorrelation functions (see also the {\it End Matter} for details). In Ref.~\cite{Mori_SymmetrizedGap_2023}, the slowest decay rate of autocorrelation functions is given by the symmetrized Liouvillian gap for quantum Markov systems. This result can be interpreted in terms of the numerical range. If we consider zero-mean observables $\mathbf{a}'$ and $\mathbf{b}'$ that satisfy $\mathbf{\pi}^\top\mathbf{a}' = \mathbf{\pi}^\top\mathbf{b}' = 0$, we can define the reduced numerical range $W_r(\tilde{R}) = \{ \langle \mathbf {x}|\tilde{R}|\mathbf {x} \rangle/\langle \mathbf {x} | \mathbf {x} \rangle | ~\mathbf {x} = \Pi^{1/2}(\mathbf{a}'+ i \mathbf{b}') \in \mathbb{C}^n , \mathbf{\pi}^\top\mathbf{a}' = \mathbf{\pi}^\top\mathbf{b}' = 0, \mathbf {x} \neq 0 \}$, which exhibits a spectral gap $g_s$. We can obtain the bound on the decay of the autocorrelation function $ {{C}_{a'a'}(t)} \leq C_{a'a'}(0) \exp \{- g_s t\}$. In fact, $W_r(\tilde{R}) (\subset W (\tilde{R}))$ is located entirely within the sectorial region, which captures essential physical information about the system. 
It may be useful to consider an appropriately defined reduced numerical range, in a similar manner, to examine a tighter bound on the spectrum, which may also imply a bound on the spectral gap.

Finally, we would like to mention an interesting topic related to Uhl and Seifert's conjecture~\cite{Uhl_conjecture_2019}. In 1938, Kolmogorov considered the collection of all eigenvalues of all $n \times n$ stochastic matrices $\Omega_n$~\cite{mashreghi2007conjecture}, and a complete description of the collection $\Omega_n$ has been discussed~\cite{romanovskydiscrete, romanovskyRecherches, dmitriev1945characteristic} and given exactly~\cite{dmitriev1946characteristic, karpelevich1951characteristic, ito1997new}. The boundary of $\Omega_n$ consists of the vertices of $k$-gons ($k\leq n$) on the Gershgorin circle (unit circle $\mathbb{S}^1 \coloneqq \{z \in \mathbb{C} \big| |z|=1 \}$) and the curves that connect them~\cite{karpelevich1951characteristic, ito1997new}. 
Since stochastic matrices are obtained from rate matrices by considering transition probabilities, this result may be reinterpreted in terms of the set $\Omega_n'$, defined as the collection of all eigenvalues of all $n\times n$ rate matrices. By  considering the relative shape of $\Omega_n$ with respect to the unit circle $\mathbb{S}^1$, one may infer the geometric boundary of $\Omega_n'$ embedded in the corresponding Gershgorin circle of radius $\alpha$. However, it is not obvious that the scaling of its relative shape provides the collection of all eigenvalues of all $n\times n$ rate matrices that are restricted by a finite thermodynamic force. Thus, we could propose the following open problem: 
Does vertical scaling by the factor $\beta'$ applied to the collection of all eigenvalues of all $n\times n$ rate matrices produce the collection of all eigenvalues of all $n\times n$ rate matrices constrained by a finite value of a thermodynamic force? 
Indeed, we find numerical evidence that supports this conjecture (see Supplementary Material \cite{supp}).
Interestingly, the opening angle in the bound by Ohga {\it et al.}~\cite{Ohga_corsscorr_2023} coincides with a part of the $n$-gon that is vertically scaled by the factor $\beta'$ on the ellipse $EL_c$. 
Thus, Uhl and Seifert's conjecture, in conjunction with the newly proposed open problem, remains an open problem and warrants further investigation.

\textit{Acknowledgments} G.-H. X. thanks Keiji Saito for fruitful discussions on the numerical range and correlation functions.
S. I. and A. K. thank Naruo Ohga for fruitful discussions on Uhl and Seifert's conjecture.
We acknowledge the support of JST ERATO Grant Number JPMJER2302, Japan. S.I. is supported by JSPS KAKENHI Grants No.~22H01141, No.~23H00467, and No.~24H00834, 
JST ERATO Grant No.~JPMJER2302, 
and UTEC-UTokyo FSI Research Grant Program.
J.-C. D. acknowledges funding from the Project INTER/FNRS/20/15074473 `The Circo' on Thermodynamics of Circuits for Computation, funded by the F.R.S.-FNRS (Belgium) and FNR (Luxembourg), and from the SIDDARTA Concerted Research Action (ARC) of the F\'{e}d\'{e}ration Wallonie-Bruxelles (Belgium). J.-C. D. is also a FY2025  JSPS fellow and Invited Professor at Department of Physics, Kyoto University (Japan). 
A.K. is supported by the European Union’s Horizon 2020 research and innovation programme under the Marie Skłodowska-Curie Grant Agreement No. 101068029 and by the John Templeton Foundation (grant 62828).

\bibliographystyle{apsrev4-2}
\bibliography{reference_bound.bib}

\newpage
\pagebreak
\clearpage

\section{End Matter}

\textit{Eigenvalues and correlation functions.---}
By definition, the numerical range $W(\tilde{R})$ is the set of all possible values of $\langle \mathbf{x} |\tilde{R}| \mathbf{x} \rangle/\langle \mathbf{x} | \mathbf{x} \rangle$. 
When $\mathbf{x} = \tilde{ \mathbf{u} }$, which is the right eigenvector of $\tilde{R}$ satisfying $\tilde{R} \tilde{ \mathbf{u} } = \lambda \tilde{ \mathbf{u} } $, we obtain $\langle \tilde{ \mathbf{u} }|\tilde{R}| \tilde{ \mathbf{u} } \rangle/\langle \tilde{ \mathbf{u} }| \tilde{ \mathbf{u} } \rangle = {\rm Re}\lambda + i (
{\rm Im} \lambda)$. Using $\tilde{R} = \Pi^{-1/2} R \,\Pi^{1/2}$, we find that $\mathbf{u} = \Pi^{1/2} \tilde{ \mathbf{u} }$ is a right eigenvector of $R$, i.e., $R \mathbf{u} = \lambda \mathbf{u}$. Following the notation in the main text, we introduce the observables such that $\tilde{ \mathbf{u} }=\Pi^{1/2}(\mathbf {a}+i\mathbf {b})$, from which it follows that
$\mathbf {a}+i\mathbf {b} = \Pi^{-1} \mathbf{u}$. Therefore, we obtain $ {\rm Re}\lambda =[\dot{C}_{aa}(0)+\dot{C}_{bb}(0) ]/[C_{aa}(0) + C_{bb}(0)]$ and ${\rm Im}\lambda =[\dot{C}_{ab}(0)-\dot{C}_{ba}(0) ]/[C_{aa}(0) + C_{bb}(0)]$ for observables $a_i = \mathrm{Re}\, u_i/\pi_i$ and $b_i = \mathrm{Im}\, u_i/\pi_i$, as discussed in Ref. \cite{Ohga_corsscorr_2023}.

\textit{Derivation of the ellipse theorem.---}
Here we provide the proof based on physical quantities. A mathematically oriented derivation is presented in Supplementary Material \cite{supp}.
First, we define the current matrix 
$
J \coloneqq R \Pi - \Pi R^{\top} = 2\Pi^{\frac{1}{2}}\mathrm{Skew}(\tilde{R})\Pi^{\frac{1}{2}}
$
with
$J_{ij} = R_{ij} \pi_{j} - R_{ji} \pi_{i}$
and the traffic matrix 
$
A \coloneqq R \Pi + \Pi R^{\top} =2 \Pi^{\frac{1}{2}}\mathrm{Sym}(\tilde{R})\Pi^{\frac{1}{2}}
$
with
$A_{ij} = R_{ij} \pi_{j} + R_{ji} \pi_{i}$.
Then, $z = x+iy\in W(\tilde{R})$ is given by
\begin{equation}
x= \frac{1}{2} \frac{a^{\top} A \,a + b^{\top} A \,b}{a^{\top}\Pi\, a + b^{\top}\Pi\, b} 
= \frac{\dot{C}_{aa}(0)+\dot{C}_{bb}(0)}{C_{aa}(0) + C_{bb}(0)}
\end{equation}
and 
\begin{equation}
y = \frac{1}{2} \frac{a^{\top} J \,b - b^{\top} J \,a}{a^{\top}\Pi\, a + b^{\top}\Pi\, b}
= \frac{\dot{C}_{ab}(0)-\dot{C}_{ba}(0) }{C_{aa}(0) + C_{bb}(0)}
.
\end{equation}
It is noted that $J$ is a zero matrix when the detailed balance condition is satisfied, which implies that $y=0$ in equilibrium systems. 

The imaginary part $y$ can be written as 
\begin{equation}
y =  \frac{1}{2} \frac{\sum_{ij} J_{ij} (a_i b_j - a_j b_i) }{\sum_{i} \pi_i(a_i^2 + b_i^2)}.
\end{equation}
The term $a_i b_j - a_j b_i$ corresponds to the cross product of the vectors $(a_i,b_i)$ and $(a_j,b_j)$, which can be written as $a_i b_j - a_j b_i = r_i r_j \sin \theta_{ij}$, where $r_i = \sqrt{a_i^2 + b_i^2}$, and $\theta_{ij}$ is the angle between the two vectors.
We then rewrite the sum over $i$, $j$ as a sum over the set $\epsilon^+$ of all edges $e=j \to i$ such that $J_e:=J_{ij} >0$.
For each edge $e$, we define $\theta_e := \theta_{ij}$ and $A_e \coloneqq A_{ij}$.
With these notations, we obtain
\begin{align}
|y| & = \frac{ \left| \sum_{e \in \epsilon^+} J_{e} r_i r_j \sin \theta_{e} \right|}{\sum_{i} \pi_i r_i^2 }  \notag \\
& \leq \sum_{e \in \epsilon^+} \frac{J_{e}}{A_{e}} \frac{ A_e r_i r_j \sin |\theta_{e}|}{\sum_{i} \pi_i r_i^2 } \notag \\
& \leq \left( \max_e \frac{J_{e}}{A_{e}} \right) \sum_{e \in \epsilon^+} \frac{ A_e r_i r_j \sin |\theta_{e}|}{\sum_{i} \pi_i r_i^2 } \notag  \\
& = \tanh \left(\max_e \frac{F_e}{2} \right) \sum_{e \in \epsilon^+} \frac{ A_e r_i r_j \sin |\theta_{e}|}{\sum_{i} \pi_i r_i^2 } \notag \\
&=: \beta \tilde{y}.
\end{align}
Here, the thermodynamic force is given by
\begin{equation}
    F_e = \ln \frac{R_{ij} \pi_j}{R_{ji} \pi_i} = \ln \frac{A_e+J_e}{A_e-J_e} =2~ \mathrm{arctanh} \left( \frac{J_e}{A_e} \right) ,
\end{equation}
which leads to ${J_{e}}/{A_{e}} = \tanh (F_e/2)$, and by the monotonicity of the function $\tanh(x)$, it follows that $\max_e ({J_{e}}/{A_{e}}) = \tanh \left(\max_e F_e /2\right) =:\beta$. 
For simplicity, we denote
\begin{equation}
h_e :=\frac{ A_e r_i r_j}{\sum_{i} \pi_i r_i^2 } > 0,
\end{equation}
and
\begin{equation}
\tilde{y} := \sum_{e \in \epsilon^+} h_e \sin |\theta_e| \geq \frac{|y|}{\beta} .
\end{equation}

The real part
\begin{equation}
x = \frac{1}{2} \frac{\sum_{i j} A_{ij} (a_i a_j+b_i b_j)}{\sum_{i} \pi_i(a_i^2 + b_i^2)} \leq 0
\end{equation}
can be decomposed into two components,
$x = x_{\rm{D}} + x_{\rm{ND}}$, corresponding to the contribution from the diagonal part of $A$, 
\begin{align}
x_{\rm{D}}
&\coloneqq \frac{1}{2} \frac{\sum_{i} A_{ii} (a_i^2+b_i^2)}{\sum_{i} \pi_i(a_i^2 + b_i^2)} \notag \\
& \geq \frac{1}{2} \min_i \frac{A_{ii}}{\pi_i} = -\max|R_{ii}| =: -\alpha,
\end{align}
and the off-diagonal part of $A$,
\begin{align}
x_{\rm{ND}}
& \coloneqq \frac{1}{2} \frac{\sum_{i \neq j} A_{ij} (a_i a_j+b_i b_j)}{\sum_{i} \pi_i(a_i^2 + b_i^2)} \notag \\
& = \sum_{e\in \epsilon^+} \frac{ A_{e} r_i r_j \cos \theta_{e}}{\sum_{i} \pi_i r_i^2 } 
=\sum_{e\in \epsilon^+} h_e \cos \theta_{e}.
\end{align}
Consider the sum of squares
\begin{align}
x_{\rm{ND}}^2 + \tilde{y}^2 
& = \left( \sum_{e\in \epsilon^+} h_e \cos \theta_{e} \right)^2 + \left( \sum_{e\in \epsilon^+} h_e \sin |\theta_{e}| \right)^2 \notag \\
& = \sum_{e_1, e_2 \in \epsilon^+} h_{e_1} h_{e_2} \cos(|\theta_{e_1}|-|\theta_{e_2}|) \notag \\
& \leq \sum_{e_1, e_2 \in \epsilon^+} h_{e_1} h_{e_2} 
= \left( \sum_{e\in \epsilon^+} h_e \right)^2,
\end{align}
and 
\begin{align}
\sum_{e\in \epsilon^+} h_e 
& = \frac{1}{2} \frac{\sum_{i \neq j} A_{ij} r_i r_j}{\sum_{i} \pi_i r_i^2 } \\
& \leq \frac{1}{2} \frac{\sum_{i \neq j} A_{ij} (r_i^2 + r_j^2)/2}{\sum_{i} \pi_i r_i^2 } \\
& = -\frac{1}{2} \frac{\sum_{i} A_{ii} r_i^2}{\sum_{i} \pi_i r_i^2 }  = x_{\rm{D}},
\end{align}
we obtain
$
x_{\rm{ND}}^2 + \tilde{y}^2 \leq x_{\rm{D}}^2,
$
which leads to
\begin{equation}
(x-x_{\rm{D}})^2 + \tilde{y}^2 \leq x_{\rm{D}}^2.
\end{equation}
This inequality implies that $(x,\tilde{y})$ is located inside a disk centered on $(x_{\rm{D}},0)$ with radius $|x_{\rm{D}}|$. Since $-\alpha<x_{\rm{D}}<0$, all these disks are enclosed by the largest one given by
$
(x+\alpha)^2 + \tilde{y}^2 \leq \alpha^2.
$
Moreover, noting that
$
{|y|}/{\beta} \leq \tilde{y} ,
$
we finally obtain
\begin{equation}
\left( \frac{x + \alpha}{\alpha} \right)^2 + 
\left( \frac{y}{\alpha \beta} \right)^2
\leq 1.  \qed
\end{equation}

\begin{figure}
    \centering
    \includegraphics[width=0.95 \columnwidth]{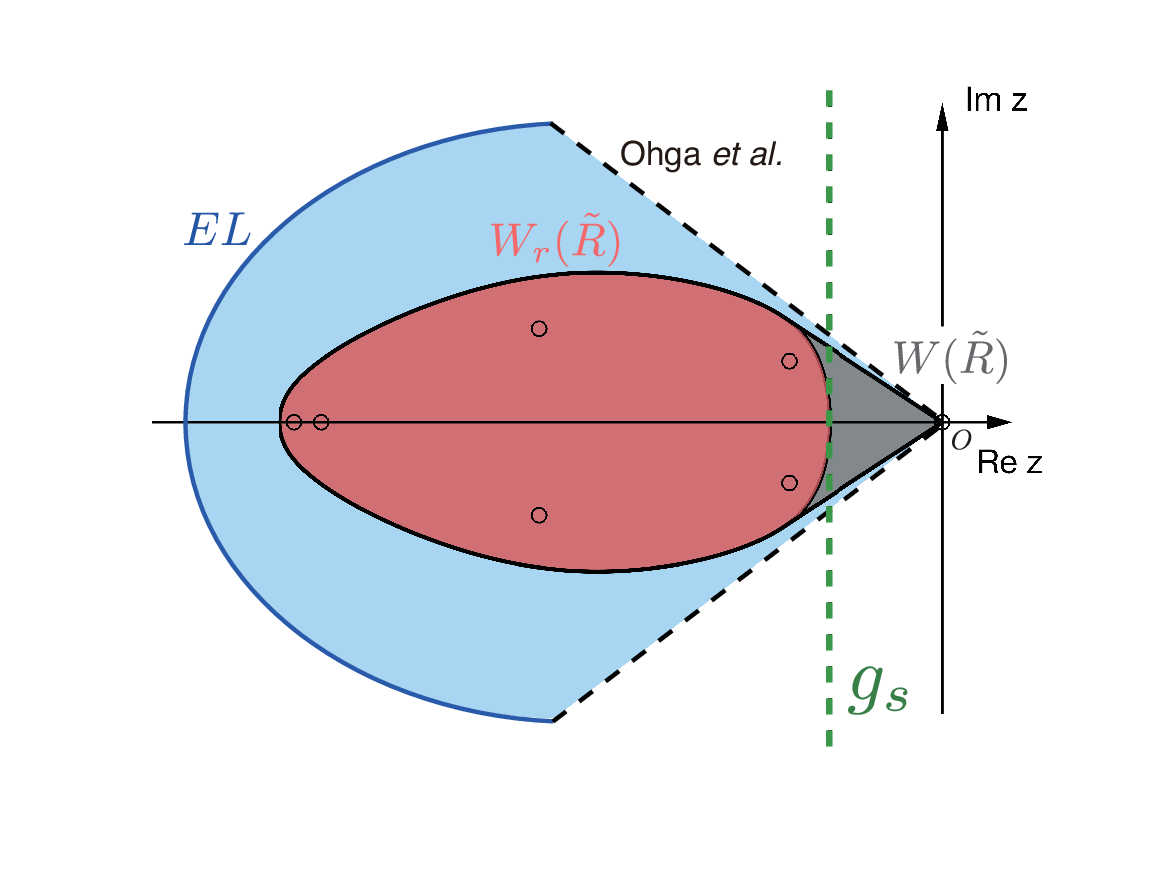}
    \caption{\RaggedRight
Reduced numerical range and spectral gap.
An illustration of the symmetrized Liouvillian gap $g_s$ (green dashed line) given by the maximum real part of the complex number in the reduced numerical range $W_r(\tilde{R})$ (red shaded region). The blue shaded region represents the thermodynamic geometric bound, and the black shaded region shows the numerical range $W(\tilde{R})$. The numerical ranges are calculated using the algorithm in Ref. \cite{Johnson_calculation_1978}.
}
    \label{fig:gap}
\end{figure}

\textit{Spectral gap and its relation to temporal correlations.}---
It has been shown in Ref.~\cite{Mori_SymmetrizedGap_2023} that, for quantum Markov systems, the slowest decay rate of autocorrelation functions is given by the symmetrized Liouvillian gap. Here, we rephrase this result in the language of the numerical range. We consider observables $\mathbf{a}'$ and $\mathbf{b}'$ with zero mean values, $\mathbf{\pi}^\top\mathbf{a}' = \mathbf{\pi}^\top\mathbf{b}' = 0$. 
By substituting the correlation functions of $\mathbf{a}'$ and $\mathbf{b}'$ into the expression of the numerical range, Eq.~\eqref{NR_corr}, we obtain a reduced numerical range $W_r(\tilde{R}) = \{ \langle \mathbf {x}|\tilde{R}|\mathbf {x} \rangle/\langle \mathbf {x} | \mathbf {x} \rangle | ~\mathbf {x} = \Pi^{1/2}(\mathbf{a}'+ i \mathbf{b}') \in \mathbb{C}^n , \mathbf{\pi}^\top\mathbf{a}' = \mathbf{\pi}^\top\mathbf{b}' = 0, \mathbf {x} \neq 0 \}$, as shown in Fig.~\ref{fig:gap}. 
Note that the real part of $W_r(\tilde{R})$ is given by 
\begin{equation}
\frac{\langle \mathbf {x}|\mathrm{Sym}(\tilde{R})|\mathbf {x} \rangle}{\langle \mathbf {x} | \mathbf {x} \rangle} < 0
\end{equation}
where the symmetrized matrix is defined as 
\begin{equation}
    \mathrm{Sym} (\tilde{R}) \coloneqq \frac{\tilde{R} + \tilde{R}^{\top}}{2}.
\end{equation}
In the subspace orthogonal to $\mathbf{\pi}$, the real part of $W_r(\tilde{R})$ is a linear combination of the nonzero eigenvalues of $\mathrm{Sym}\{\tilde{R}\}$. As a result, $W_r(\tilde{R})$ exhibits a spectral gap
\begin{equation}
-g_s = \max_{|\mathbf {x} \rangle \neq 0: \langle \mathbf {\pi}|\mathbf {x} \rangle = 0} \frac{\langle \mathbf {x}|\mathrm{Sym}(\tilde{R} )|\mathbf {x} \rangle}{\langle \mathbf {x} | \mathbf {x} \rangle},
\end{equation}
which is the second largest eigenvalue of $\mathrm{Sym}(\tilde{R})$ and coincides with the symmetrized Liouvillian gap introduced in Ref.~\cite{Mori_SymmetrizedGap_2023}. 
Since $|\mathbf{x} \rangle$ can be chosen as eigenvectors corresponding to nonzero eigenvalues of $\tilde{R}$, and $\tilde{R}$ is a similarity transformation of $R$, all nonzero eigenvalues of $R$ lie within $W_r(\tilde{R})$. It therefore follows that 
\begin{equation}
    g \geq g_s,
\end{equation}
where $g (>0)$ is the spectral gap of $R$. 

Moreover, based on the definition of the numerical range $W(\tilde{R})$ and following the method in Ref.~\cite{Mori_SymmetrizedGap_2023}, we can obtain the bound on the decay of the autocorrelation function.
Given that 
$C_{a'a'}(t) \coloneqq \mathbf {a}'^{\top}e^{Rt} \Pi\, \mathbf {a}'$ with $\mathbf {\pi}^{\top} \mathbf {a}' = 0$, we can define $\mathbf {a}'_t \coloneqq e^{R^{\top}t} \mathbf {a}'$  with $\mathbf {\pi}^{\top} \mathbf {a}'_t = 0$. According to the Cauchy-Schwarz inequality, we obtain
\begin{equation}
    \left (\frac{C_{a'a'}(t)}{C_{a'a'}(0)}  \right )^2 \leq \frac{{\mathbf {a}'_t}^{\top} \Pi\, \mathbf {a}'_t}{{\mathbf {a}'}^{\top} \Pi \, \mathbf {a}'} .
\end{equation}
The time derivative of ${\mathbf {a}'_t}^{\top} \Pi \, \mathbf {a}'_t$
is given by
\begin{align*}
\frac{\mathrm{d}}{\mathrm{d}t}( {\mathbf {a}'_t}^{\top} \Pi \, \mathbf {a}'_t )
& = {\mathbf {a}'_t}^{\top} (R\Pi + \Pi R^{\top}) \mathbf {a}_t \\
&= 2 \langle \mathbf {x}'|\mathrm{Sym}\{\tilde{R}\} | \mathbf {x}' \rangle  \\
& \leq -2 g_s \langle \mathbf {x}' | \mathbf {x}' \rangle \\
&= -2 g_s {\mathbf {a}'_t}^{\top} \Pi \, \mathbf {a}'_t,
\end{align*}
where we denote $\mathbf {x}' = \Pi^{1/2} \, \mathbf{a}'_t$, and the inequality is obtained from the definition of the spectral gap $g_s$.
Thus, ${\mathbf {a}'_t}^{\top} \Pi \mathbf {a}'_t \leq \mathbf {a}'^{\top} \Pi \, \mathbf {a}' ~ \exp \{- 2 g_s t\}$.
Finally, we obtain $ {{C}_{a'a'}(t)} \leq C_{a'a'}(0) \exp \{- g_s t\}$.

The numerical range $W(\tilde{R})$ is the convex hull of the reduced numerical range $W_r(\tilde{R})$ and the origin, satisfying the relation $W_r(\tilde{R}) \subset W(\tilde{R}) \subseteq W({R})$. 
Moreover, $W_r(\tilde{R})$ is located entirely within the sectorial region, which provides a tight convex set that captures essential physical information about the system.
In this sense, the matrix $\tilde{R}$ is referred to as a sectorial matrix, which has been the subject of recent interest in matrix analysis and operator theory \cite{ARLINSKII2003133,WANG2020152,WANG2023441,ALAKHRASS2021179,DRURY2024108,LI2025173}.

\newpage
\pagebreak
\clearpage 
\onecolumngrid

\begin{center}	
        \textbf{\large --- Supplementary Material --- }
\end{center}

\setcounter{equation}{0}
\setcounter{figure}{0}
\setcounter{table}{0}
\setcounter{page}{1}
\makeatletter
\renewcommand{\theequation}{S\arabic{equation}}
\renewcommand{\thefigure}{S\arabic{figure}}
\renewcommand{\bibnumfmt}[1]{[S#1]}
\renewcommand{\citenumfont}[1]{S#1}
\renewcommand{\bibliography}

\subsection{An alternative proof of the theorem}

A more mathematically oriented derivation is presented in the following.
To prove the ellipse theorem, we introduce the matrix
\begin{equation*}
\tilde{S} \coloneqq \frac{1}{2} (\tilde{R} + \tilde{R}^{\top}) + \frac{1}{2 \beta } (\tilde{R} - \tilde{R}^{\top}).
\end{equation*}
Observe that $\tilde{S}_{ii} = \tilde{R}_{ii}$ for all $i$. In addition, from the definition of $F_e$, we have 
\begin{equation*}
    \beta = \tanh \left(\max_e \frac{F_e}{2} \right) = \max_{i,j} \left ( \frac{\tilde{R}_{ij} -\tilde{R}_{ji}}{\tilde{R}_{ij} + \tilde{R}_{ji}} \right),
\end{equation*}
which ensures that $\tilde{S}_{ij} \geq 0$ for all $i\neq j$. 
It can be verified that $S\coloneqq \Pi^{1/2} \tilde{S} \Pi^{-1/2}$ is a valid rate matrix with stationary distribution $\mathbf{\pi}$.
Here, $\tilde{S}$  is constructed by rescaling the antisymmetric part of $\tilde{R}$ by a factor, and $\beta$ is the minimal factor that ensures $\tilde{S}_{ij} \geq 0$, which guarantees that $S$ is a valid rate matrix.

The matrix 
\begin{equation*}
    I + \frac{\tilde{S}}{\alpha}
\end{equation*}
with $\alpha \coloneqq \max_i |R_{ii}| = \max_i |\tilde{S}_{ii}|$ has all nonnegative entries. Therefore, its numerical radius 
\begin{equation*}
    r := \max \{ |z|, z\in W(I + \tilde{S} / \alpha)  \}
\end{equation*} 
is equal to the largest eigenvalue of 
\begin{equation*}
    I + \frac{\tilde{S} + \tilde{S}^{\top}}{2\alpha} ,
\end{equation*}
as stated in Theorem 2.1 of Ref.~\cite{Goldberg}.
Observe that $I + (\tilde{S} + \tilde{S}^{\top} )/ (2\alpha)$ is related by a similarity transform to 
\begin{equation*}
    I + \frac{S + \Pi S^{\top} \Pi^{-1} }{2\alpha} ,
\end{equation*}
which is a valid stochastic matrix. From the Perron-Frobenius theorem, its largest eigenvalue is $1$. Therefore, the numerical radius $r$ of $I + \tilde{S} / \alpha$ is 1, which implies that its numerical range falls within a unit disk centered at the origin. 
As a consequence, the numerical range of $\tilde{S}$ falls within a disk of radius $\alpha$ centered at $-\alpha$,
\begin{equation}
    |z+\alpha|\leq \alpha \text{~~~~~~ for all ~~~~~~} z \in W(\tilde{S}). \label{circle}
\end{equation}

Finally, we may express $\tilde{R}$ a convex combination of $\tilde{S}$ and its transpose:
\begin{equation*}
    \tilde{R} = (1-\eta)\tilde{S} + \eta\tilde{S}^{\top},
\end{equation*}
where $\eta = (1-\beta)/2$. This transformation is known as Levinger’s transformation, and it implies that the numerical range of $\tilde{R}$ is related to the numerical range of $\tilde{S}$ by a rescaling of the imaginary axis [Eq. (3), Ref.~\cite{Maroulas}],
\begin{equation*}
W(\tilde{R}) = \left\{ {\rm Re} \, z + i(1 - 2\eta ) \, {\rm Im}\, z \,\middle|\, z \in W(\tilde{S}) \right\}.
\end{equation*}
Since $1-2\eta = \beta$, combining with the circular bound in Eq.~\eqref{circle} implies that the numerical range of $W(\tilde{R})$ falls within an ellipse of width $2\alpha$ and height $2\alpha \beta$ centered at $-\alpha$ on the real axis. 

\subsection{Conjecture on the boundary of eigenvalues of rate matrices}

We begin by presenting the Karpelevich theorem for stochastic matrices \cite{Karpelevich,Ito}. 
Let $P \in \mathbb{R}^{n \times n}$ be an $n$-dimensional stochastic matrix, where $P_{ij} > 0$ and $\sum_i P_{ij} = 1$ for all $j$.
The set $\Omega_n$ of all eigenvalues of $P$ lies within the unit disc in the complex plane.
The Karpelevich theorem characterizes the precise structure of the boundary of $\Omega_n$.

\textbf{Theorem} \cite{Karpelevich,Ito}.
The region \( \Omega_n \) is contained within the unit disk \( |z| \leq 1 \), and intersects the unit circle \( |z| = 1 \) at the points \( e^{2\pi i \frac{a}{b}} \), where \( a \) and \( b \) are relatively prime integers satisfying \( 0 \leq a \leq b \leq n \). 
The boundary of \( \Omega_n \) consists of these intersection points and a collection of curvilinear arcs connecting them in circular order. Let the endpoints of one such arc be \( e^{2\pi i \frac{a_1}{b_1}} \) and \( e^{2\pi i \frac{a_2}{b_2}} \), with \( b_1 \leq b_2 \). Each arc is described by the following parametric equation:
\begin{equation}
    \lambda^{b_2} (\lambda^{b_1} - t)^{\left\lfloor \frac{n}{b_1} \right\rfloor} 
    = (1 - t)^{\left\lfloor \frac{n}{b_1} \right\rfloor} \lambda^{b_1 \left\lfloor \frac{n}{b_1} \right\rfloor},
\end{equation}
where \( t \in [0, 1] \) is a real parameter.
Here, \( \left\lfloor \cdot \right\rfloor \) denotes the floor function, i.e., the greatest integer less than or equal to the given number.

Now, we consider the set \( \Omega_n' \) of eigenvalues of the Markov rate matrix \( R \). 
According to the conjecture of Uhl and Seifert \cite{Uhl_Seifert}, the region \( \Omega_n' \) is assumed to lie within the ellipse
\begin{equation}
EL_c \coloneqq \left\{ z \in \mathbb{C} \,\Bigg|\, \left( \frac{\mathrm{Re}\,z + \alpha}{\alpha} \right)^2 + 
\left( \frac{\mathrm{Im}\,z}{\alpha \beta'} \right)^2 \leq 1 \right\}, \notag
\end{equation}
where \( \alpha = \max_i |R_{ii}| \) and \( \beta' = \tanh \left( \max_c \left\{ \mathcal{F}_c / (2n_c) \right\} \right) \).

\textbf{Conjecture.}  
The region \( \Omega_n' \) of eigenvalues of the Markov rate matrix \( R \) is the affine image of \( \Omega_n \) under the transformation:
\begin{equation}
    \Omega_n' = \left\{ \alpha(\mathrm{Re}\,z - 1) + i \alpha \beta'\,\mathrm{Im}\,z \,\middle|\, z \in \Omega_n \right\}. \label{Karpelevich_rate}
\end{equation}

\begin{figure}[t]
  \centering

  \begin{subfigure} [b]{0.46\textwidth}
  \includegraphics[width=1\textwidth]{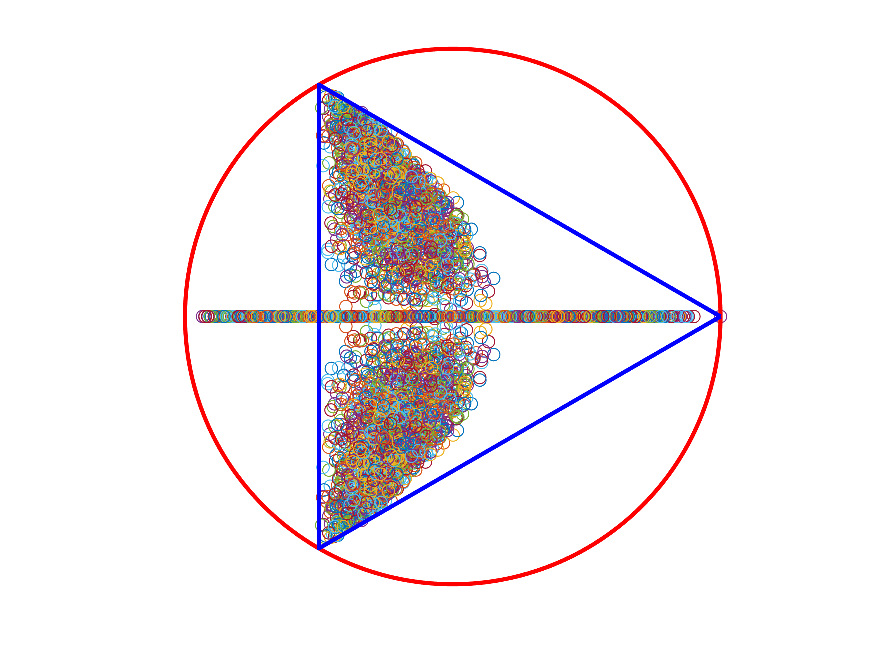}
    \caption{\centering $n=3$}
    \label{fig:sub1}
  \end{subfigure}
  \hspace{5pt}
  \begin{subfigure}[b]{0.46\textwidth}
    \includegraphics[width=1\textwidth]{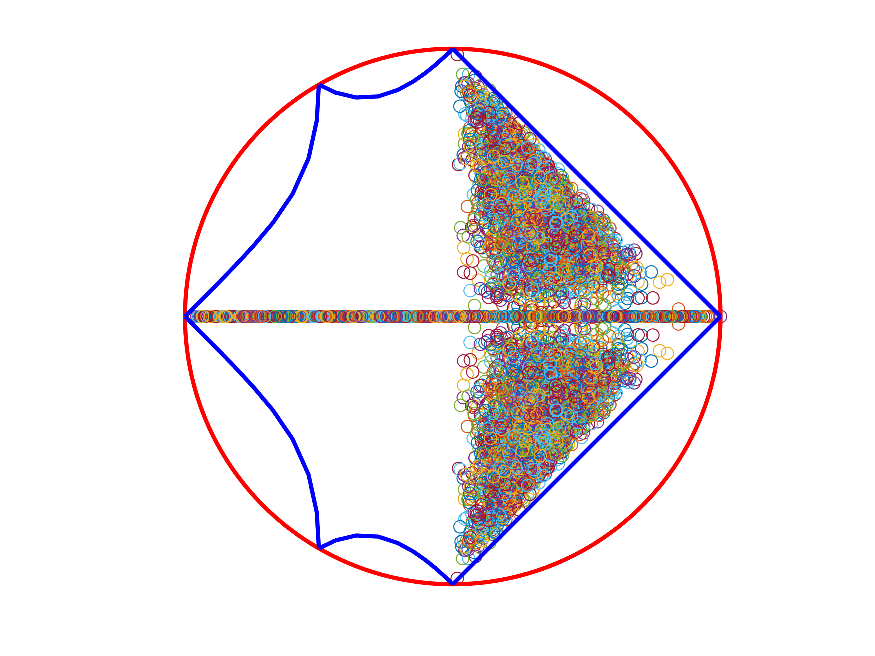}
    \caption{\centering $n=4$}
    \label{fig:sub2}
  \end{subfigure}
  
  \vspace{5pt}

  \begin{subfigure}[b]{0.46\textwidth}
    \includegraphics[width=1\textwidth]{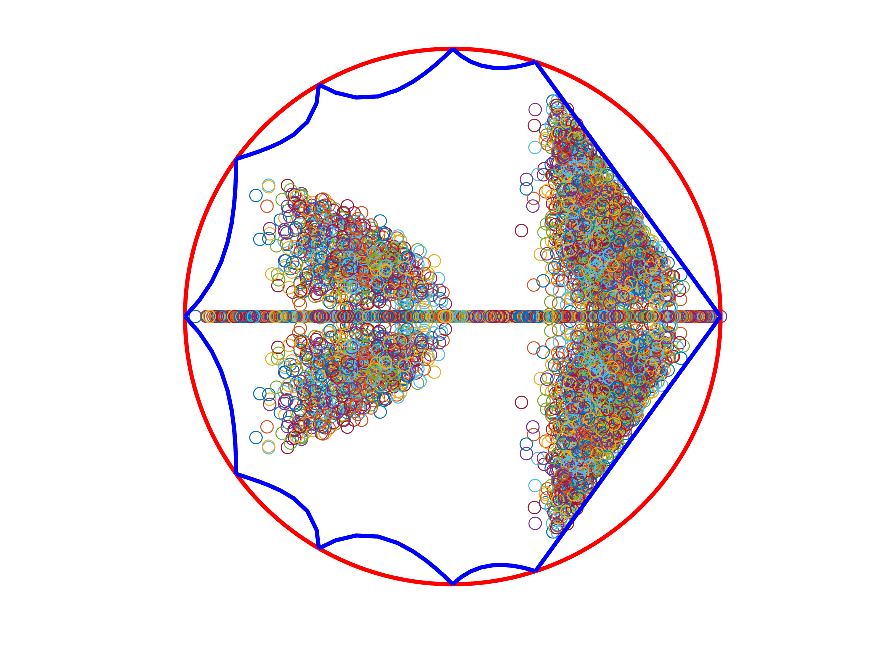}
    \caption{\centering $n=5$}
    \label{fig:sub3}
  \end{subfigure}
  \hspace{5pt}
  \begin{subfigure}[b]{0.46\textwidth}
    \includegraphics[width=1\textwidth]{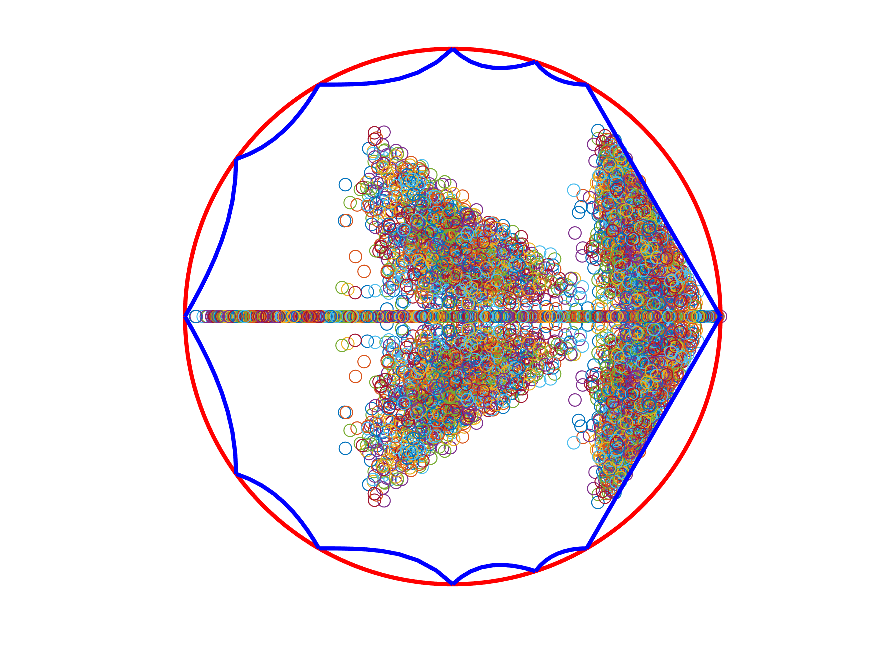}
    \caption{\centering $n=6$}
    \label{fig:sub4}
  \end{subfigure}
  
  \vspace{5pt}

  \begin{subfigure}[b]{0.46\textwidth}
    \includegraphics[width=1\textwidth]{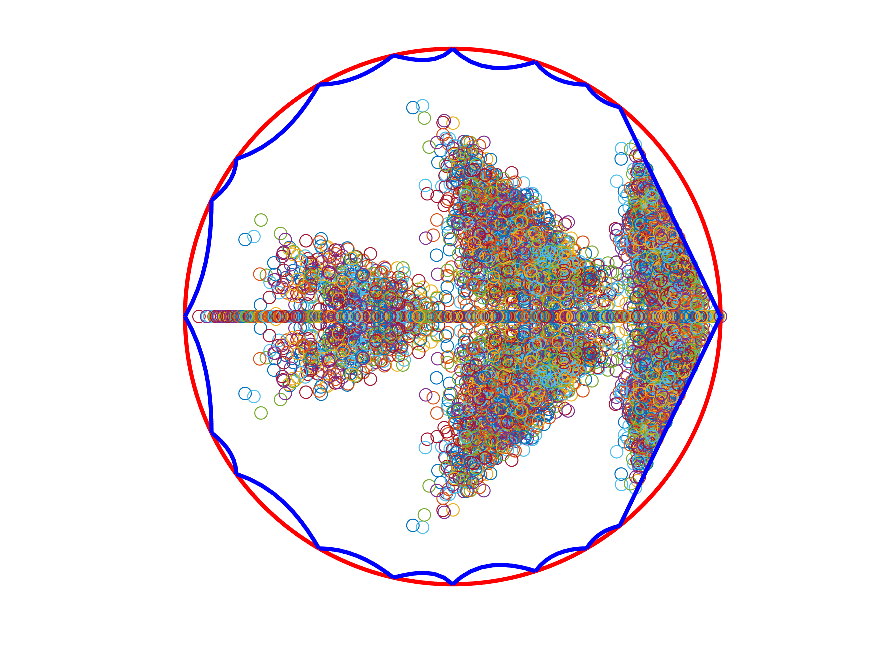}
    \caption{\centering $n=7$}
    \label{fig:sub5}
  \end{subfigure}
  \hspace{5pt}
  \begin{subfigure}[b]{0.46\textwidth}
    \includegraphics[width=1\textwidth]{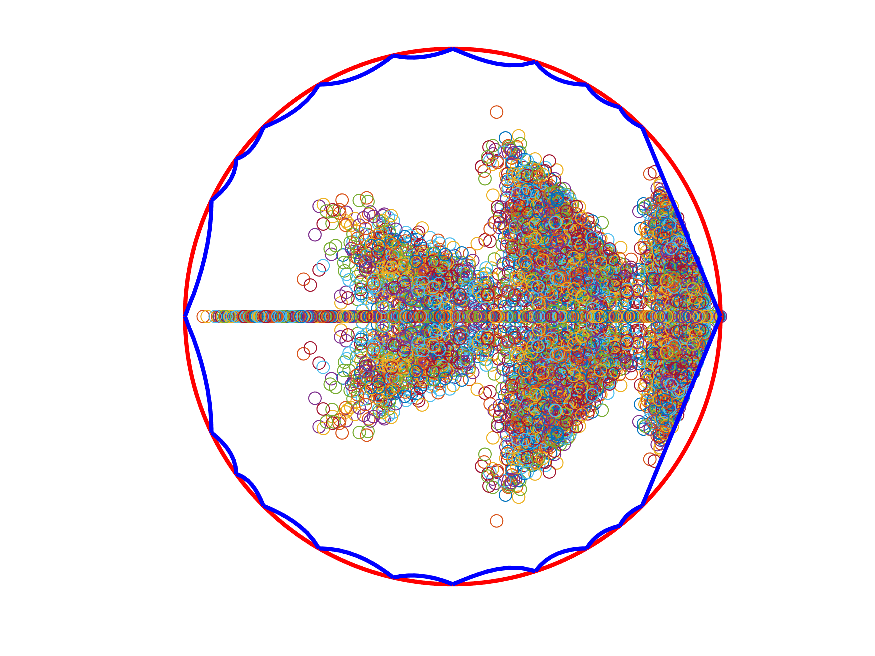}
    \caption{\centering $n=8$}
    \label{fig:sub6}
  \end{subfigure}

  \caption{\RaggedRight
  Numerical results for the eigenvalue spectra of rate matrices with various dimensions \( n \). 
  We consider only single-cycle cases with random transition rates and illustrate the relative shape of $\Omega_n'$ with respect to Uhl and Seifert's ellipse by mapping $\Omega_n'$ to $\Omega_n$ using the inverse of the transformation in Eq.~\eqref{Karpelevich_rate}. The unit circle is shown in red, the boundary of $\Omega_n$ is shown in blue, and the transformed spectra are shown by hollow circles.}
  \label{fig:eigen_spectra}
\end{figure}

As shown in Fig.~\ref{fig:eigen_spectra}, we consider a Markov jump process on a single cycle with random transition rates. The plot is generated using the inverse of the transformation in Eq.~\eqref{Karpelevich_rate}, which shows the relative shape of $\Omega_n'$ with respect to the ellipse proposed by Uhl and Seifert. Since we are only considering a single cycle, its shape differs from that of $\Omega'_n$. However, we can confirm that all the eigenvalues are in $\Omega'_n$, and there are no apparent counterexamples to our conjecture in this simple case. We note that the eigenvalues of an $n' \times n'$ matrix become the eigenvalues of an $n \times n$ matrix ($n\geq n')$. Thus, the union set with cases less than or equal to $n$ is closer to the form of $\Omega'_n$.
We also note that the eigenvalues corresponding to the points $e^{2\pi i \frac{a}{n}}$ exist for an integer $a (\leq n)$ when we consider the uniform transition rates~\cite{Uhl_Seifert}. Therefore,  the eigenvalues corresponding to the points $e^{2\pi i \frac{a}{n'}}$ also exist for integers $a$ and $n'$ satisfying $a \leq n' \leq n$. However, random matrices are not suitable for generating uniform transition rates. 

Interestingly, the arcs connected to the origin appear as straight lines, which coincide with the sectorial bound established by Ohga \textit{et al.}~\cite{Ohga_Supplementary}.

\end{document}